\documentclass[preprint,12pt]{elsarticle}

\usepackage{amssymb}
\usepackage{rotating}
\usepackage{amssymb}
\usepackage{amsmath}
\usepackage{graphicx}
\usepackage{algorithm}
\usepackage{algorithmic}
\usepackage{subfigure}
\usepackage{epsfig}
\usepackage{comment}
\usepackage{url}
\usepackage{float}
\usepackage{slashbox}
\usepackage{enumerate}

\journal{EPL}

\begin{document}

\begin{frontmatter}



\title{Community Structure Detection in Complex Networks with Partial Background Information}

\author[label1]{Zhong-Yuan Zhang}
\address[label1]{School of Statistics, Central University of Finance and Economics, \\ P.R.China}



\begin{abstract}
Constrained clustering has been well-studied in the unsupervised learning society. However, how to encode constraints into community structure detection, within complex networks, remains a challenging problem. In this paper, we propose a semi-supervised learning framework for community structure detection. This framework implicitly encodes the \emph{must-link} and \emph{cannot-link} constraints by modifying the adjacency matrix of network, which can also be regarded as de-noising the consensus matrix of community structures. Our proposed method gives consideration to both the topology and the functions (background information) of complex network, which enhances the interpretability of the results. The comparisons performed on both the synthetic benchmarks and the real-world networks show that the proposed framework can significantly improve the community detection performance with few constraints, which makes it an attractive methodology in the analysis of complex networks.
\end{abstract}



\end{frontmatter}


\section{Introduction}
\label{intro}
Evidences have shown that there are often modules or community structures in complex networks \cite{Girvan02}. For example, a community could be a set of proteins that have similar functions in a protein-protein interaction (PPI) network, or it could be a group of fans that like visiting similar kind of music web pages, or a university club, etc.  Though there is still no standard and clear definition of community structure, we may regard a community in complex networks as a set of nodes that have similar link-pattern, or in other words, these nodes have similar preference and connect to the other nodes in a similar way. The most common and widely studied community is a subgraph that is densely interconnected but loosely connected with the rest of the graph. Meanwhile, there are also other types of communities. Discovering communities is very important for revealing the organization and the functions of the network, such as understanding how the units in some systems communicate with each other and work together, or learning how the new ideas or diseases spread in a group of persons \cite{spread}, etc.

How to detect community structures has thus become a hot topic, and many interesting models and algorithms have been developed and have achieved good results. But all of these methods are in essence a kind of unsupervised learning, meaning that they only make use of the network topology information. However, in many real scenarios, there is usually some background information that could also be used in detecting the communities. This information can be treated as additional constraints, and how to combine the information with the network topology to guide the detecting process is an interesting problem that is worthy of working on.

In this paper, we propose a semi-supervised framework to incorporate prior information into community structure detection. Our framework is flexible to integrate various known information. One can easily provide pairwise constraints on a few nodes in the network, specifying whether they must or cannot be in the same community structure, based on the background information and domain knowledge. For example, the nodes that have similar functions should be must-link, or the nodes that have different opinions should be cannot-link. The framework implicitly encodes the must-link and cannot-link constraints by modifying the adjacency matrix of the network, which can also be regarded as the de-noising process of the consensus matrix of the community structures, i.e., creating connections within communities and
removing connections across communities.

\section{Semi-supervised learning for community structure detection}
\label{semi0}
In this section, we formulate our semi-supervised framework for community structure detection. Firstly,
we introduce the definition of adjacency matrix $A^{[0]}$ of an undirected and unweighted simple graph $G$ with $n$ nodes:
$$A^{[0]}_{ij} = \left\{\begin{array}{rcl}
         1, & & \mbox{if}\ i\sim j\\
         0, & & \mbox{if}\ i = j\ \mbox{or}\ i\nsim j,\\
      \end{array}\right.
$$
where $i\sim j$ means there is an edge between node $i$ and $j$, and $i\nsim j$ means there is no edge between them. Here $A^{[0]}$ is $n\times n$ and symmetric.

Note that the diagonal elements of $A^{[0]}$ are all zeros, but these zeros are obviously different from the ones at the off-diagonal positions which mean there are no connections between the nodes. Hence we here set the diagonal elements of $A^{[0]}$ to 1. The revised adjacency matrix is denoted by $A^{[1]}$. Another variation of $A^{[0]}$ is its complementary matrix $C^{[A]}=1-A^{[0]}.$
\subsection{Incorporating prior knowledge into adjacency matrix}\label{semi}
In many real applications, we often have some background information that can be used for community structure detection. Specifically, we consider the following two types of pairwise constraints:
\begin{itemize}
\item \texttt{Must-Link constraints} $C_{ML}$: $(i,j)\in C_{ML}$ means that the two nodes $i$ and $j$ must belong to the same community,
\item \texttt{Cannot-Link constraints} $C_{CL}$: $(i,j)\in C_{CL}$ means that the two nodes $i$ and $j$ cannot belong to the same community.
\end{itemize}
We incorporate the constraints $C_{ML}$ and $C_{CL}$ into the adjacency matrix 
$A^{[1]}$
to get a new matrix $B$ as follows:
\begin{equation}\label{eq:02}
B_{ij} = \left\{\begin{array}{rcl}
         \alpha, & & \mbox{if}\, (i,j)\in C_{ML}\\
         0, & & \mbox{if}\, (i,j)\in C_{CL}\\
         A^{[1]}_{ij} & & \mbox{otherwise},
      \end{array}\right.
\end{equation}
where $\alpha$ is a positive constant.

As one can see, if we set $\alpha$ to 1, and for all the pairs of nodes, we know whether they should belong to $C_{ML}$ or $C_{CL}$, or in other words, we know very well the community structures in the graph, the adjacency matrix will reduce to the standard consensus matrix, whose $(i,j)th$ element means whether node $i$ and node $j$ are in the same community, $1$ means yes and $0$ means no. Hence from the viewpoint of consensus matrix, incorporating prior knowledge can be regarded as the de-noising process.

We have tried different $\alpha$, i.e., $\alpha=1$ and $\alpha=2$, and the results of $\alpha=2$ always get better. We omit the comparisons here due to space limit.

After incorporating background information into the adjacency matrix, we then apply nonnegative matrix factorization (NMF), spectral clustering and InfoMap, which are of the most common and widely-used models in unsupervised learning, for community structure detection.
\subsection{Nonnegative matrix factorization (NMF, \cite{nmf99, nmf00,Chen,bayesian})}
\label{NMF}
NMF can be expressed as follows: given a nonnegative objective matrix $X$ of size $n\times m,$ columns of which are samples and rows are features, we try to find two nonnegative matrices $F$ of size $n\times k$ and $G$ of size $m\times k$ such that: $X\approx FG^T.$ This problem is often formulated as the following nonlinear programming:
\begin{eqnarray}\label{eq:01}
\min_{F, G} & &  J(X\|FG^T)\\\nonumber
      s.t.           & & F\geqslant0, G\geqslant0,
\end{eqnarray}
where $J(X\|FG^T)$ is the cost function that measures the dissimilarity between $X$ and $FG^T$, and $\geqslant0$ means that $F$ and $G$ should not have negative entries. The most popular algorithm designed for
NMF is multiplicative update rules.
The objective matrix $X$ for NMF can be selected as $B$.

In \cite{wang2008clustering}, it showed that the diffusion-kernel-based similarity matrix $SK$\footnote{Definition of diffusion kernel $K$ and the similarity matrix $SK$ \cite{wang2008clustering, diffusionkernel}: $K=\lim\limits_{n\to\infty}(1+\displaystyle\frac{\beta L}{n})^n=expm(\beta L)$, where $L$ is the opposite Laplacian of $A^{[0]}$: $$
    L_{ij}=\left\{
            \begin{array}{rcl}\vspace{1mm}
             1 & & \mbox{if}\ i\sim j\\\vspace{1mm}
             -d_i & & \mbox{if}\ i = j\\
             0 & & \mbox{otherwise,}
            \end{array}
            \right.
$$
and $d_i$ is the degree of node $i$. $$SK_{ij}=\frac{K_{ij}}{\sqrt{K_{ii}K_{jj}}}.$$
  We set $\beta = 0.2$ in this paper.
Note that there is a MATLAB command ``expm'' for the exponential of a matrix.} was the best choice for the objective matrix $X$ among all the candidates, hence we also tested the performance of $SK$ in this paper.

The community structures of the network can be obtained from $G$: node $i$ is of community $k$ if $G_{ik}$ is the largest element in the $i$th row of $G$.
\begin{enumerate}[1)]
\item Standard NMF with least squares error: If $J(X\|FG^T)$ is selected as the least squares error: $J(X\|FG^T)=\|X-FG^T\|_F^2$, the algorithm of multiplicative update rules can be summarized in Algorithm \ref{Al:01}.
In this paper, the iteration number \texttt{iter} is set to 100.
\hspace*{50mm}
\begin{algorithm}
\caption{Nonnegative Matrix Factorization
 (Least Squares Error)}
\label{Al:01}
\begin{algorithmic}[1]
\REQUIRE $X,$ iter
\ENSURE $F, G.$
\FOR{$t=1:\mbox{iter}$}
\STATE \vspace{3mm}
$
\displaystyle F_{ik}:=F_{ik}\frac{(XG)_{ik}}{(FG^{T}G)_{ik}}
$
\STATE \vspace{3mm}
$
\displaystyle G_{ik}:=G_{ik}\frac{(X^TF)_{ik}}{(GF^{T}F)_{ik}}
$
\ENDFOR
\end{algorithmic}
\end{algorithm}

\item Standard NMF with K-L divergence: If $J(X\|FG^T)$ is selected as the KL divergence:
$\displaystyle J(X\|FG^T)=\sum_{i,j}[X_{ij}\log\frac{X_{ij}}{(FG^T)_{ij}}-X_{ij}+(FG^T)_{ij}],$ the corresponding update rules of $F$ and $G$ are:
$$\displaystyle F_{ik}:=\frac{F_{ik}}{\sum_jG_{jk}}\sum_j\frac{X_{ij}}{(FG)_{ij}}G_{jk};
$$
$$\displaystyle G_{jk}:=\frac{G_{jk}}{\sum_iF_{ik}}\sum_i\frac{X_{ij}}{(FG)_{ij}}F_{ik}.
$$
\item Symmetric NMF (SNMF): There is a variant of NMF for semi-supervised clustering, whose objective function can be formulated as: $\|X-GSG^T\|_F^2$. The update rules of $G$ and $S$ are \cite{Chen}:
    $$
    G_{ik}:=G_{ik}\frac{(XGS)_{ik}}{(GSG^TGS)_{ik}};
    $$
    $$
    S_{ik}:=S_{ik}\frac{(G^TXG)_{ik}}{(G^TGSG^TG)_{ik}}.
    $$
\item Bayesian NMF \cite{bayesian}: It optimizes the NMF model under the Bayesian framework, and can get better results under some circumstances.
\end{enumerate}
\subsection{Spectral Clustering \cite{ng2002spectral}}\label{spectral}
Spectral clustering is very powerful in its simplicity and effectiveness, which can be summarized in Algorithm \ref{Al:02}. Note that there are many variations of the standard one, and the detailed analysis can be found in \cite{spectral1, spectral2}.
\subsection{InfoMap \cite{infomap}} This model grew out of information theory, and tries to reveal the communities by optimizing a quality function about the minimum description length of random walks on the network. The model is among the best for community detection \cite{comparison}.
\begin{algorithm}
\caption{Spectral Clustering}
\label{Al:02}
\begin{algorithmic}[1]
\REQUIRE $B\in\mathbb{R}^{n\times n}$
\ENSURE Community Label $Y\in\mathbb{R}^{n\times 1}$ of the $n$ nodes
\STATE $L = D^{1/2}BD^{1/2}$, where $D$ is the diagonal matrix with the element $D_{ii}=\sum_jB_{ij}$ .
\STATE Formimg the matrix $X=[x_1, x_2, \cdots, x_k]\in\mathbb{R}^{n\times k}$, where $x_i, i=1,2,\cdots,k$ are the top $k$ eigenvectors of $L$.
\STATE Normalizing $X$ so that rows of $X$ have the same $L_2$ norm: $X_{ij}=X_{ij}/(\sum_jX_{ij}^2)^{1/2}.$
\STATE Clustering rows of $X$ into $k$ clusters by K-means.
\STATE $Y_{i}=j$ if the $i$th row is assigned to cluster $j$.
\end{algorithmic}
\end{algorithm}
\subsection{An illustrative example}
\label{example}
\begin{figure}[H]
\centering
\includegraphics[height=40mm,width=100mm]{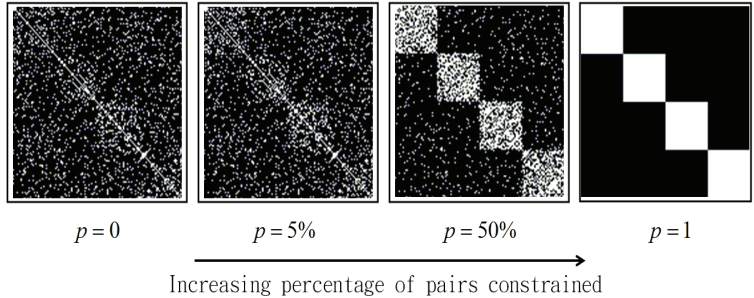}
 \caption{An illustrative example to show the process of incorporating prior information into the adjacency matrix as de-noising the consensus matrix.}\label{Fig:01}
\end{figure}

We close this section by an illustrative example as follows: we try to detect the community structures in a GN network with 128 nodes (For details, see \emph{Data Description}, $Zout=10.$).
The network has 4 communities with 32 nodes each. The heatmap of the corresponding adjacency matrix $A^{[1]}$ is shown as the leftmost in Fig. \ref{Fig:01}. If we have prior knowledge about the network structure so that we can determine a percentage of pairs of nodes as must-link or cannot-link, we can incorporate them into $A^{[1]}$. As one can see in Fig. \ref{Fig:01}, the adjacency matrix becomes more and more clear as the percentage of pairs constrained increases, and finally reduces to the standard consensus matrix of the community structures. This example demonstrates that background information is valuable to improve the accuracy of community structure detection.
\section{Experimental Results}
\label{results}
In this section, we empirically demonstrated the effectiveness of our proposed semi-supervised framework for community structure detection by applying NMF, spectral clustering and InfoMap with the de-noised consensus matrices to several well-studied networks.
\subsection{Data Description}
\label{synthetic}
We used both synthetic and real-world networks to test the effectiveness of our methods. The details of these datasets are as follows:
\begin{enumerate}[1)]
\item GN 
\cite{Girvan02}
: Maybe the most widely used benchmarks are GN networks. The network has $128$ nodes which are divided into four non-overlapping communities with $32$ nodes each. The degree of each node is $Zin +Zout=16$, in other words, each node averagely has exactly $16$ edges which randomly connect $Zin$ nodes in its own community and $Zout$ nodes in other communities. As one can see, with the increasing $Zout$, the community structures will become less clear and the problem more challenging. In this paper, we set $Zout$ to 8.
\item LFR
\cite{LFR}
: Indeed, in most of the real applications, the community structures are more complicated than GN networks. The size of the network might be larger, or the numbers
of the nodes in different communities might not be identical, or different nodes might have
different positions, i.e., some are ¡°superstars¡± or ¡°hubs¡± and should have higher degrees while the others are leaves. The LFR benchmark networks are thus proposed to address these problems. In LFR networks, both the degree and the community size distributions are power laws, with exponents $\gamma$
and $\beta$, which is more practical. Each node has a fraction $1-\mu$ of its links with the nodes in its own community and a fraction $\mu$ with the other ones. Here $\mu$ is called the mixing parameter.

We set the parameters of the LFR network as follows: the number of nodes was $1000$, the average degree of the nodes was $20$, the maximum degree was $50$, the exponent of the degree distribution $\gamma$ was 2 and that of the community size distribution $\beta$ was 1, and the mixing parameter $\mu$ was $0.8$. The communities were non-overlapping.
\item Karate \cite{zachary1977}: this dataset contains the network of friendships between 34 members of a karate club
at an American university. This club was by chance split into two smaller ones due to the divergence of
opinions about the club fees.
\item  Football \cite{Girvan02}: this dataset contains the network of American football games (not soccer) between
Division IA colleges during regular season Fall 2000. There are 115 nodes representing the football teams
while an edge means there was a game between the teams connected by the edge. The teams were divided
into 12 conferences, and all teams except few (mainly in two conferences) played against the ones in the same conference more frequently than those in other conferences.
\end{enumerate}
\subsection{Assess Standards}
\label{measure}
In our experiments, the normalized mutual information (NMI, \cite{strehl2003cluster}) was used as the standard to evaluate the community structure detection
performance. The value can be formulated as follows:

$$NMI(M_1, M_2) = \frac{\sum\limits_{i=1}^{k}
     \sum\limits_{j=1}^{k}n_{ij}\log\displaystyle\frac{n_{ij}n}{n_i^{(1)}n_j^{(2)}}}
  {\sqrt{\left(\sum\limits_{i=1}^{k}n_i^{(1)}\log\displaystyle\frac{n_i^{(1)}}{n}\right)
  \left(\sum\limits_{j=1}^{k}n_j^{(2)}\log\displaystyle\frac{n_j^{(2)}}{n}\right)}},
$$
where $M_1$ is the ground-truth cluster label and $M_2$ is the computed cluster label, $k$ is the community number, $n$ is the number of nodes, $n_{ij}$ is the number of nodes in the ground-truth cluster $i$ that are assigned to the computed cluster $j$, $n_i^{(1)}$ is the number of nodes in the ground-truth cluster $i$ and $n_j^{(2)}$ is the number of nodes in the computed cluster $j$, $\log$ is the natural logarithm.

Compared with simply counting the number of misclassified nodes, NMI is more informative, especially suitable for imbalanced datasets (i.e., the numbers of the nodes in different communities are not identical). For example, in a four-sample toy data, the ground-truth cluster label could be $1,1,1,2$. The computed cluster labels of two different models were $1, 1, 1, 1$ and $1, 1, 2, 2$ respectively. In other words, the smaller cluster was masked and not detected by the first model, hence the second model should be better though it also had one sample mis-clustered. But the accuracy (number of misclassified nodes divided by the number of nodes in the graph) results of these two models were all $75\%$, which was misleading. On the other hand, the NMI under this case was $0$ (the numerator of NMI was: $3\log\displaystyle\frac{3\cdot 4}{3\cdot 4}+1\cdot\log\displaystyle\frac{1\cdot 4}{1\cdot 4}=0$) and $34.56\%$ respectively,  which was relatively more reasonable and informative.

In the case study, we also used the modularity function $Q$ \cite{newman2004finding, newman2006modularity} as the standard to determine the best community number $k$. The function can be defined as follows:

$$
  Q=\sum_{C_k}\left[\frac{L(V_{C_k}, V_{C_k})}{L(V,V)}-\left(\frac{L(V_{C_k},V)}{L(V,V)}\right)^2\right],
$$
where $C_k$ is the $k$th community in the graph, $L(V_1,V_2)=\sum\limits_{i\in V_1, j\in V_2, i\neq j}a_{ij},$ and $a_{ij}$ is the element of $A^{[0]}.$

The larger the values of NMI and Q, the better the graph partitioning results.

Firstly, we compared the clustering performance of NMF-based models with different similarity measures including $A^{[0]}$, $A^{[1]}$, $C^{[A]}$ and $SK$.  
The results show that $A^{[1]}$ is a competitive one, though there is no single winner. Note that calculating the diffusion kernel is time consuming for large scale networks, hence we used $A^{[1]}$ for the NMF-based models in the following experiments. The details are omitted here due to space limit.
\subsection{Results Analysis}
\label{simulation}
\begin{figure}[H]
\centering
\setcounter{subfigure}{0}
\hspace{-4mm}
\subfigure[GN, $Zout=8,$ $\alpha=2.$]{\includegraphics[height=40mm,width=55mm]{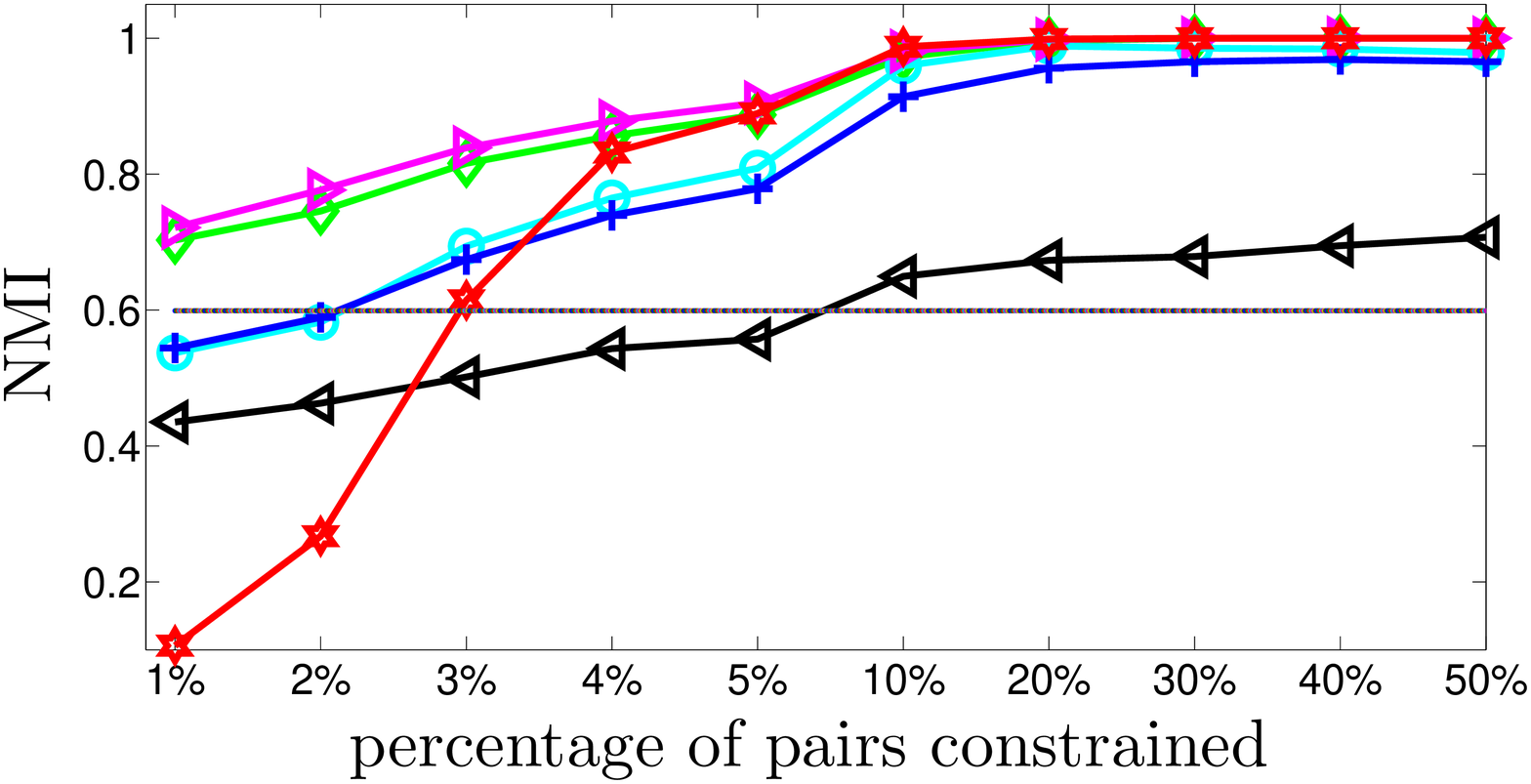}}
\hspace{5mm}
\subfigure[LFR, $\mu=0.8,$ $\alpha=2.$]{\includegraphics[height=40mm,width=60mm]{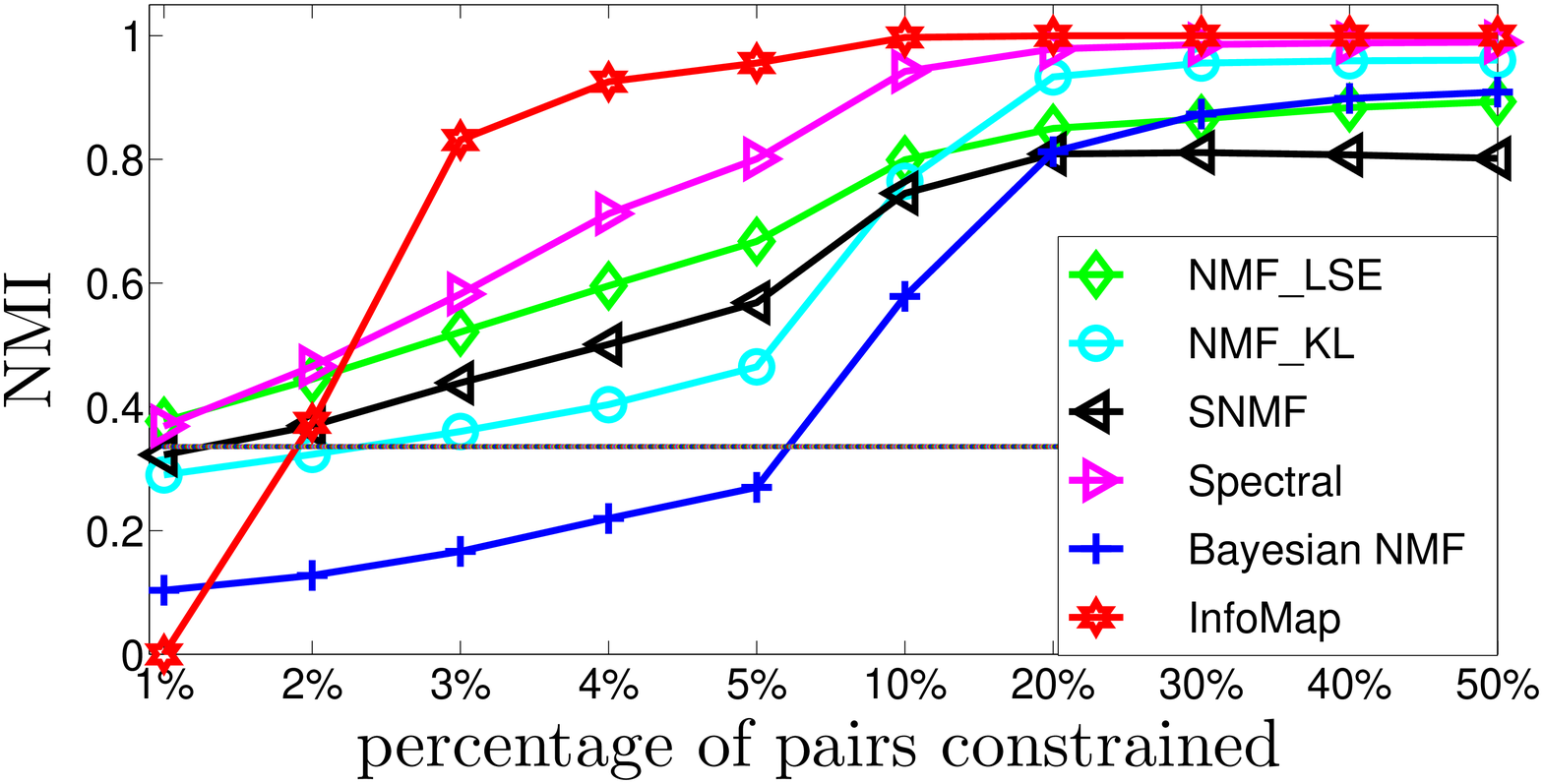}}
\caption{Averaged NMI of different models for different percentages of node pairs constrained on GN and LFR datasets. The black horizontal line is the best NMI result that had ever got by NMF\_LSE, NMF\_KL, SNMF, Bayesian NMF, spectral clustering and InfoMap with no prior knowledge available.  ``LSE'' means least squares error, ``KL'' means K-L divergence, ``SNMF'' means symmetric NMF.}
\label{Fig:04}
\end{figure}

In this subsection, we systematically compared the results of NMI obtained by the models on the artificial datasets and the karate network with prior knowledge available. For an undirected network with $n$ nodes, there are totally $n(n-1)/2$ node pairs available. We randomly picked out some pairs of nodes, and determined whether they belonged to $C_{ML}$ or $C_{CL}$: if the two nodes had the same community label, they were must-link, otherwise, they were cannot-link. The
results were averages of ten trails and given in Fig. \ref{Fig:04} and Table \ref{Tab:01}. From
these figure and table, one can observe that: i) The trends of all the models are generally
identical and the values of the averaged NMI increase with the increasing percentage of pairs constrained; ii) for synthetic datasets: GN and LFR, the model of InfoMap and the spectral clustering are better than the NMF-based models, especially for the LFR datasets; iii) for the karate network, NMF with least squares error performs better; iv) our proposed framework is flexible and model independent, or in other words, it can be naturally combined with many models, such as NMF, spectral clustering, InfoMap, etc.

In summary, our proposed semi-supervised framework does greatly enhance the results of community structure detection by benefitting from the user provided background information.

\subsection{A Case Study: College Football Network}
\label{case}
In this subsection, we used the college football network for a case study, and saw the partitioning results of NMF\_LSE given different percentages of pairs constrained. Actually, we also tried spectral clustering and got similar results. Details of spectral clustering are omitted here due to space limit.

The teams were separated into 12 conferences, and most of them played against the ones in the same conference more frequently. However, the teams 37, 43, 81, 83, 91 (in conference \emph{IA Independents}), 12, 25, 51, 60, 64, 70, 98 (in conference \emph{Sunbelt}), 111, 29 and 59 played more frequently against the ones in other conferences. Table \ref{Tab:03} lists the basic information about these teams, from which one can observe that three out of five teams in \emph{IA Independents} never played against the ones in the same conference and the other two teams played only once.

Firstly, we tried to determine the community number $k$. We compared the values of modularity $Q$ at different $k$, and the function achieved its peak value at $k=11$. By combining the results of $Q$ values in Table \ref{Tab:02} with the information in Table \ref{Tab:03}, we set the community number $k=11$ and the teams in \emph{IA Independents} would be assigned to the other eleven conferences based on the outputs of NMF. Hence there were $115-5=110$ teams with ground-truth conference labels and totally $110\times(110-1)/2=5995$ team pairs available. We randomly selected some pairs as constraints: if the two teams of the pair were in the same conference, they were must-link (ML), otherwise, they were cannot-link (CL).
\begin{table}[H]
\caption{Averaged NMI of different models given different percentages of node pairs constrained on the karate dataset.  ``P'' means percentage of node pairs constrained. Meanings of ``LSE'', ``KL'' and ``SNMF'' are identical with that in Fig. \ref{Fig:04}, and ``SP'' means spectral clustering.}
\centering
\begin{tabular}{c | c c c c}\hline\hline
 \backslashbox{P}{Models}  & NMF$\_$LSE & NMF$\_$KL & SNMF & SP \\\hline
 $1\%$ & \textbf{99.84\%} & 73.38\% & 59.53\% & 90.19\% \\
 $2\%$ & \textbf{98.86\%} & 73.44\% & 51.50\% & 90.19\%\\
 $3\%$ & \textbf{99.67\%} & 82.86\% & 54.06\% & 95.10\%\\
 $4\%$ & \textbf{99.84\%} & 85.18\% & 60.96\% & 96.73\%\\
 $5\%$ & \textbf{99.84\%} & 89.24\% & 53.74\% & 95.10\%\\
 $10\%$& \textbf{100\%} & 89.14\%   & 57.91\% & \textbf{100\%}\\
 $20\%$& \textbf{100\%} & 98.37\%   & 56.57\% & \textbf{100\%}\\\hline\hline
\end{tabular}
\label{Tab:01}
\end{table}

\begin{figure}
\setcounter{subfigure}{0}
 \subfigure[]{\includegraphics[height=45mm,width=60mm]{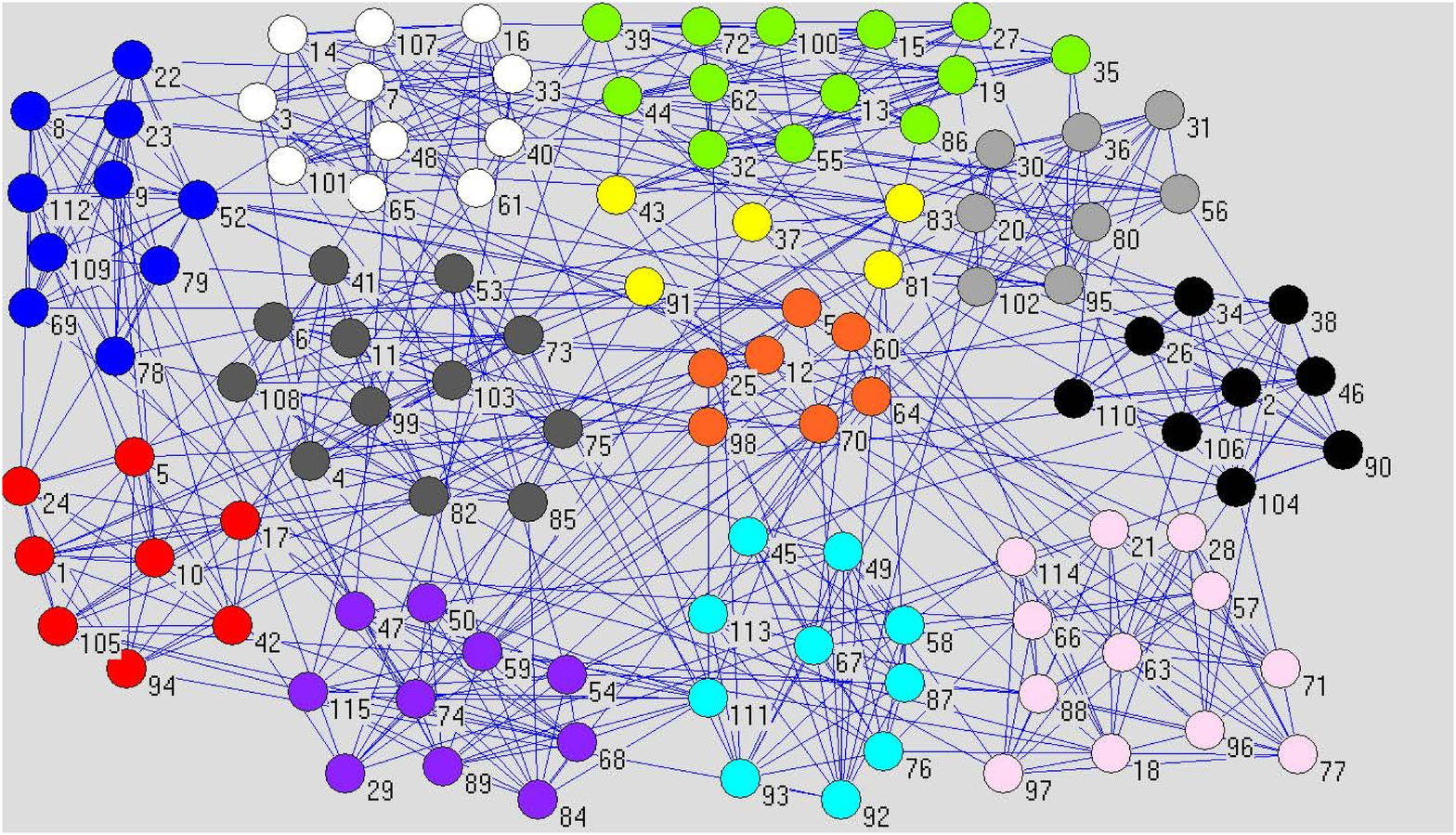}}
\setcounter{subfigure}{2} \subfigure[]{\includegraphics[height=45mm,width=60mm]{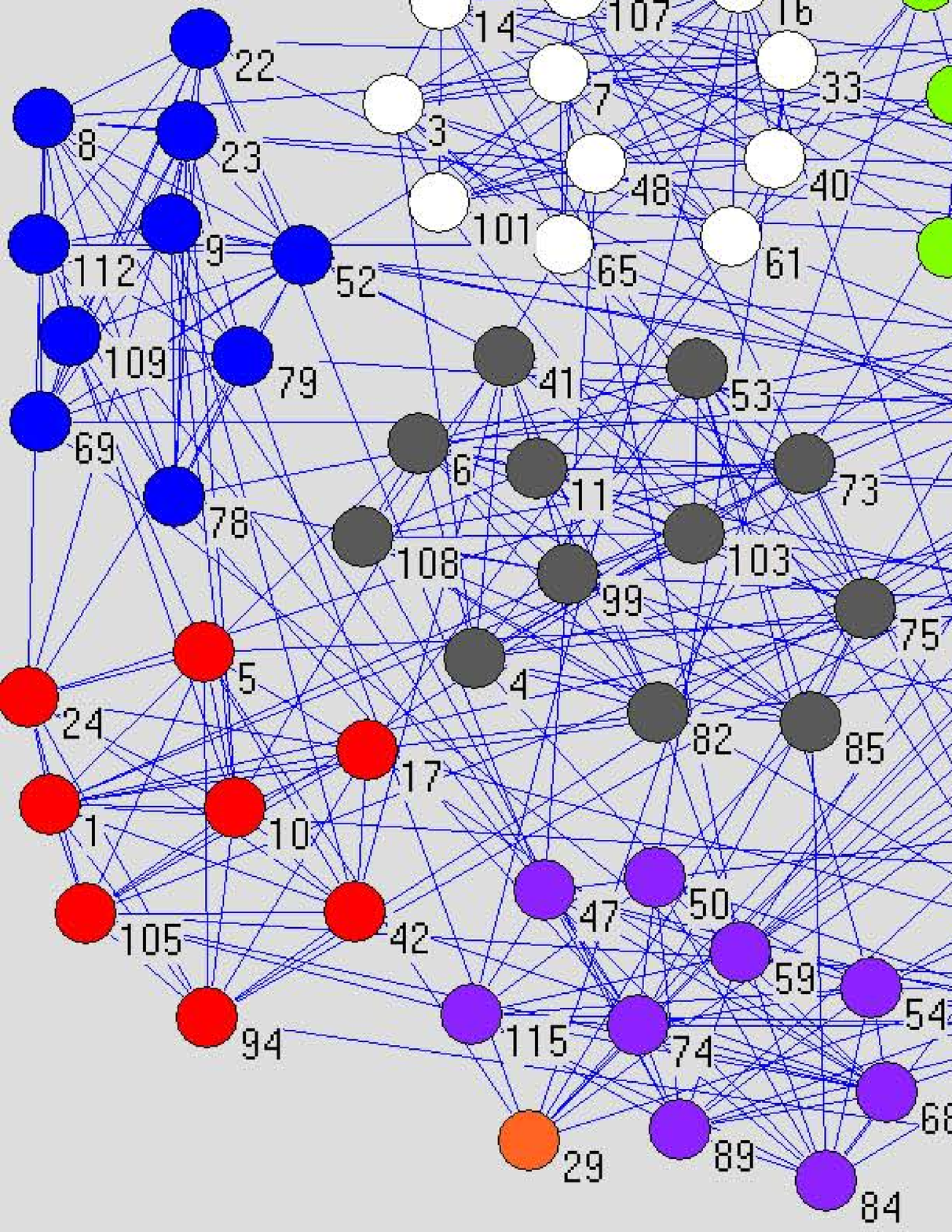}}
\setcounter{subfigure}{1}  \subfigure[]{\includegraphics[height=45mm,width=60mm]{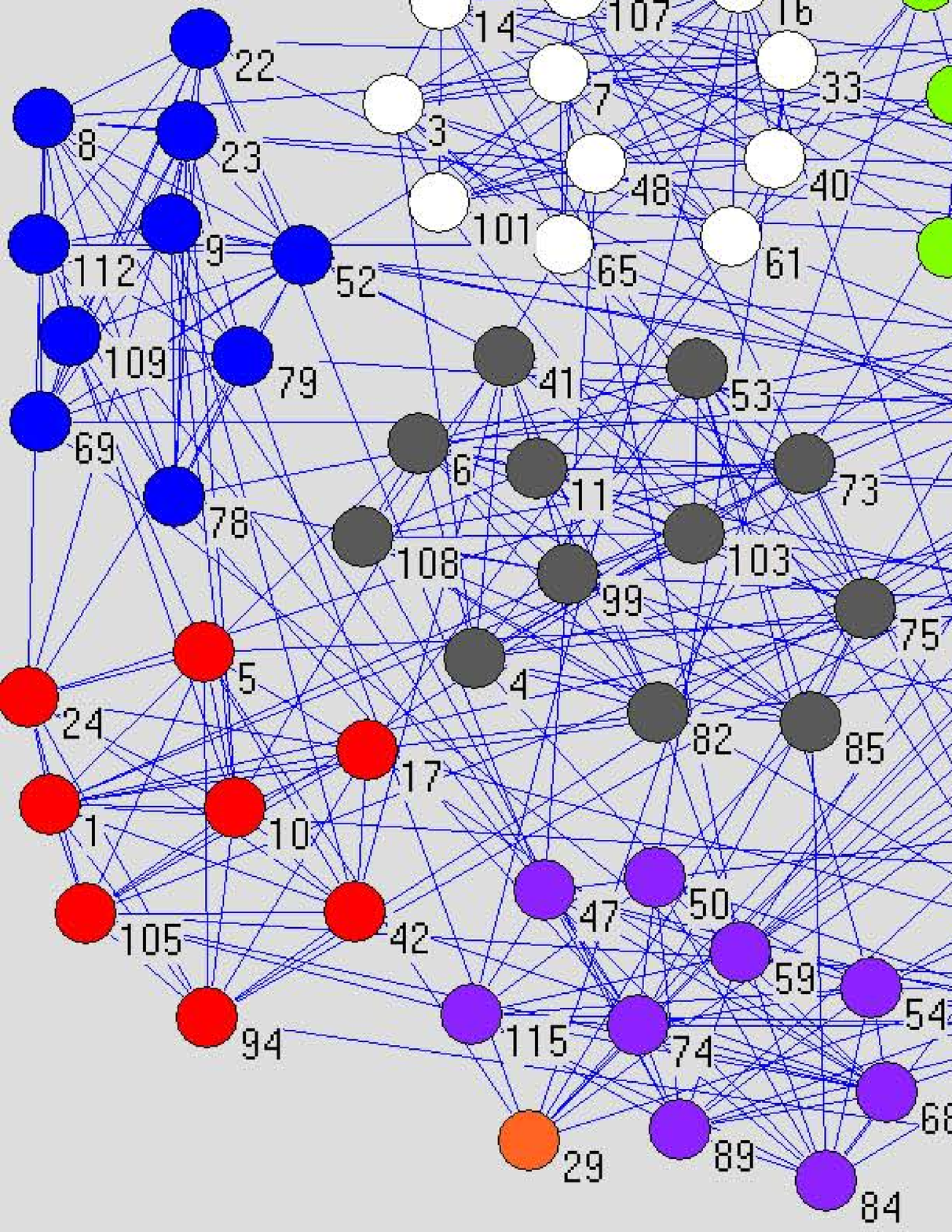}}
\setcounter{subfigure}{3} \hspace*{13mm}  \subfigure[]{\includegraphics[height=45mm,width=60mm]{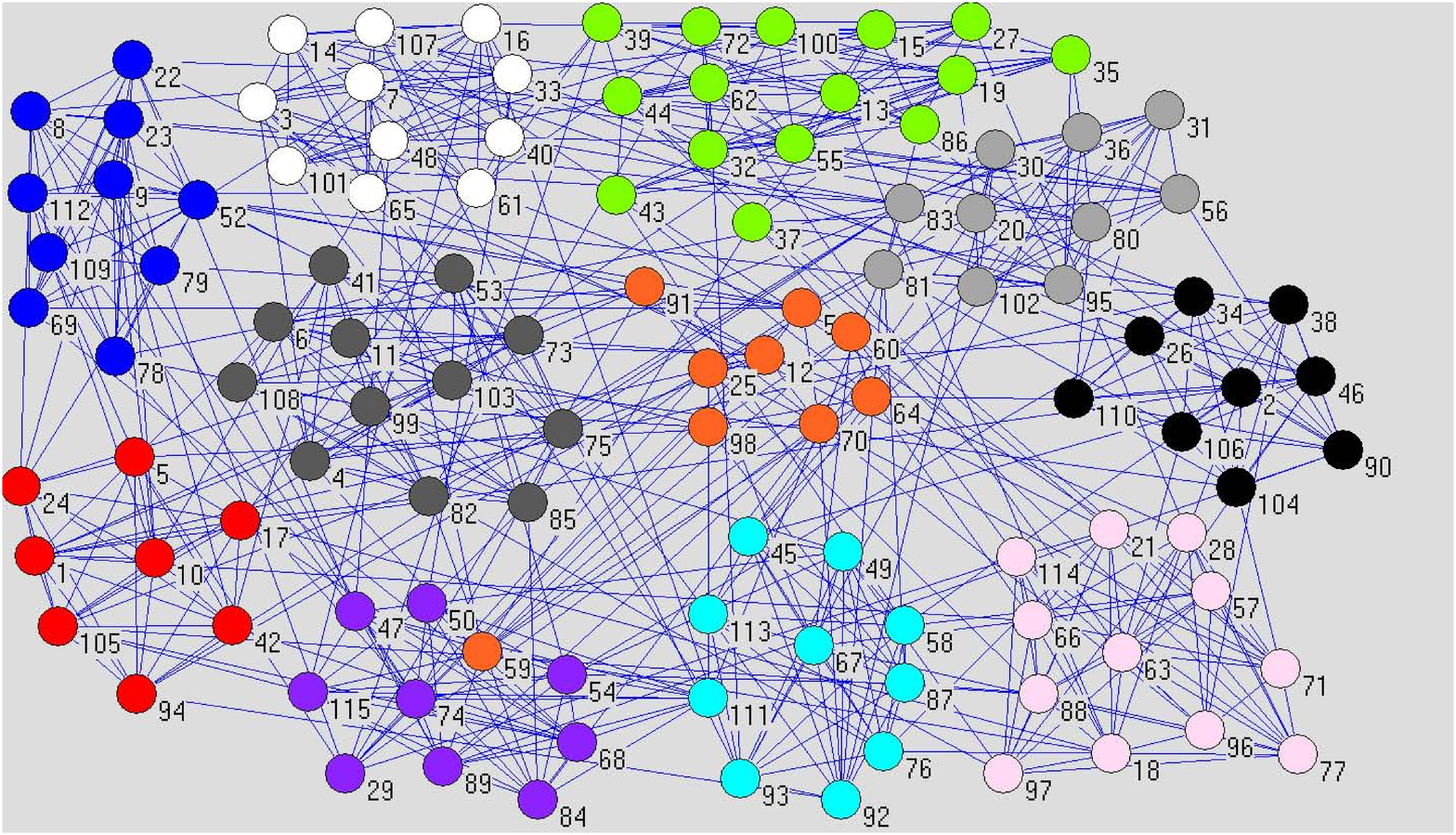}}
 \caption{Comparison of the results of NMF corresponding to different percentages of pairs constrained. (a): Real grouping in football dataset. There are 12 conferences of 8-12 teams (nodes) each. (b): Outputs of NMF without any prior knowledge. (c): Outputs of NMF corresponding to $5$ percent of pairs constrained. (d): Outputs of NMF corresponding to $20$ percent of pairs constrained.}\label{Fig:06}
\end{figure}

Figure \ref{Fig:06} gives the resulting partitions of NMF corresponding to different percentages of pairs constrained. When given no prior knowledge constrained, there were $5$ abnormal teams mis-clustered: teams 29, 60, 64, 98, 111; But after randomly given $5$ percent of pairs constrained, the results were significantly improved and only two abnormal  teams were mis-clustered: teams 29 and 111. Finally, when given $20$ percent, there was only one team mis-clustered: team 59. From these results, one can see that: 1) NMF is really good enough in that only some abnormal teams are not correctly clustered; 2) our semi-supervised clustering framework does take the background information and domain knowledge into consideration, which makes the partitioning results more explainable.
\begin{table}[H]
\caption{Basic information about the abnormal teams that played more frequently against the ones in the other conferences. ``T'' means the team id, ``F'' means the times that the team played against the other ones in the same conference or in the other conferences, ``S'' means the same conference, ``O'' means the other conferences.}
\centering
\begin{tabular}{c | c c || c | c c}\hline\hline
 \backslashbox{T}{F}  & S & O & \backslashbox{T}{F} & S & O \\\hline
 \emph{37} & 0 & 8 & \emph{60}& 2 & 6 \\
 \emph{43} & 0 & 7 & \emph{64}& 2 & 7 \\
 \emph{81} & 1 & 10 &\emph{70}& 3 & 8 \\
 \emph{83} & 1 & 10 &\emph{98}& 3 & 5 \\
 \emph{91} & 0 & 9 & \emph{111}&0 & 11\\
 \emph{12} & 4 & 6 & \emph{29}& 0 & 9 \\
 \emph{25} & 3 & 7 & \emph{59}& 2 & 8 \\
 \emph{51} & 3 & 6 &           &   & \\
 \hline\hline
\end{tabular}\label{Tab:03}
\end{table}

\begin{table}[H]
\caption{Values of averaged Q functions of NMF$\_$LSE and spectral clustering. The range of the community
number $k$ that we have tried is $8\sim 12$. The peak values were achieved at $k=11$.  Meaning of ``LSE'' is identical with that in Fig. \ref{Fig:04}.}
\centering
\begin{tabular}{c | c c}\hline\hline
 \backslashbox{Community Number}{Models}  & NMF$\_$LSE & Spectral Clustering\\\hline
 8 & 0.5770 & 0.5932  \\
 9 & 0.5831 & 0.5927 \\
 10& 0.5890 & 0.5942\\
 11& \textbf{0.5934} & \textbf{0.5978}\\
 12& 0.5885 & 0.5951\\\hline\hline
\end{tabular}\label{Tab:02}
\end{table}

\subsection{How to give the prior knowledge: randomly or based-on-rule}
\label{reinforcement}
\begin{figure}
\centering
\includegraphics[height=40mm,width=60mm]{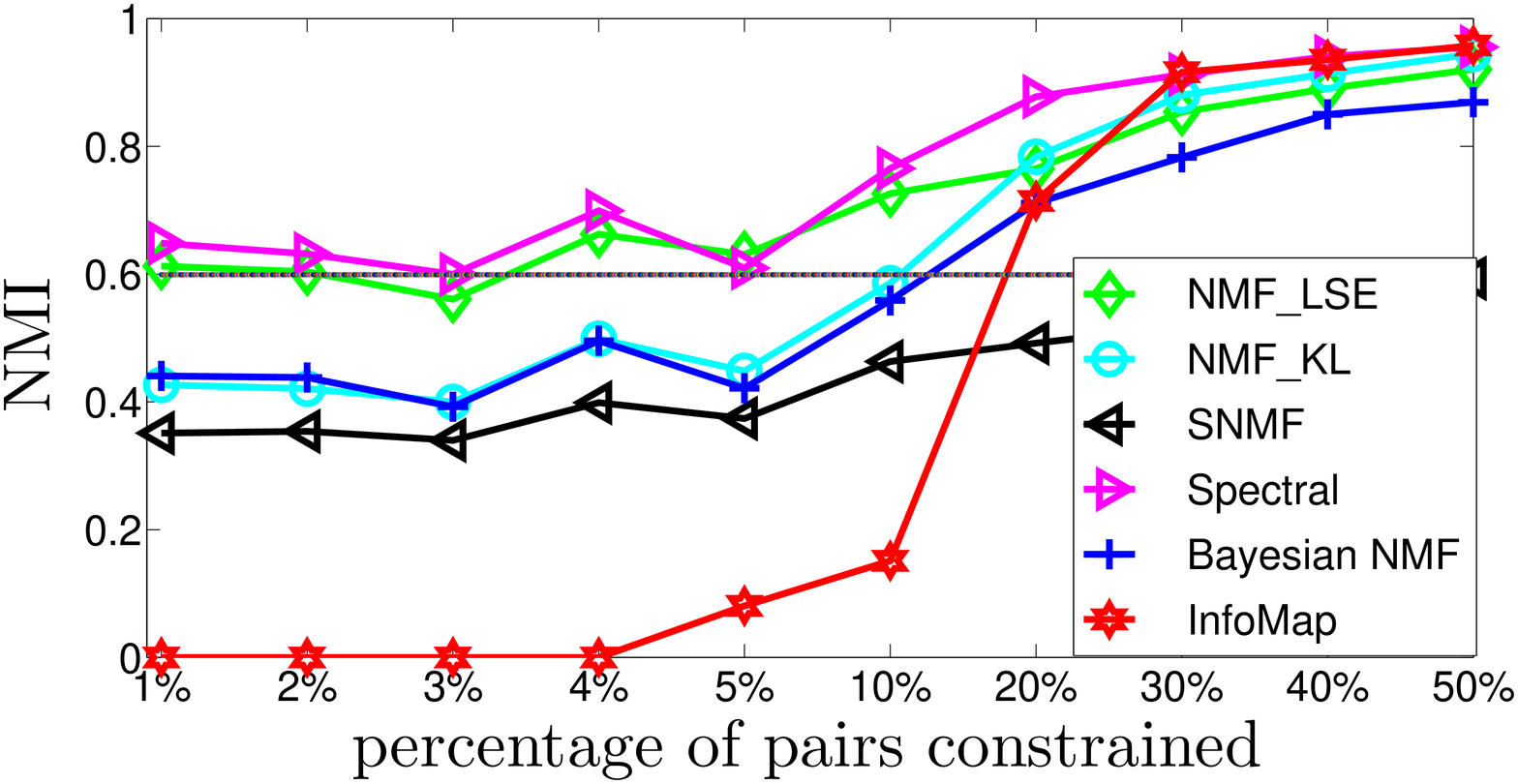}
\caption{Averaged NMI of different models for different percentages of node pairs constrained on GN datasets ($Zout=8$). The prior knowledge are given based on rule. Meanings of the black horizontal line, ``LSE'', ``KL'' and ``SNMF'' are identical with that in Fig. \ref{Fig:04}.}
\label{Fig:05}
\end{figure}

Finally, we ask an interesting question: how to select the prior information and incorporate them into the models? To the best of our knowledge, in practice, the most widely used method is to randomly select some pairs of samples or nodes and manually determine whether they are must-link or cannot-link based on the domain knowledge. But are there any better methods to select the pairs that can either reduce the workload or improve the clustering performance, or both? Indeed, for a large scale network, a very small percentage of pairs may still mean a huge workload. In this subsection, we attempted to introduce a new rule-based method to address this problem. Firstly, we computed the hamming distances between all pairs of nodes (rows of $A^{[1]}$), and sorted the distances to find the largest and the smallest ones (this step can be finished by programming calculation, not manually). We selected the pairs that have the largest distances and the smallest distances simultaneously. For example, if we wanted to select $P$ pairs of nodes, we selected $P/2$ pairs with the largest distances, and also selected $P/2$ pairs with the smallest distances. Then we manually decided whether the selected pairs were must-link or cannot-link and incorporated them into the clustering process. The results on GN datasets are shown in Fig. \ref{Fig:05}, from which one can observe that our preliminary results are not good enough compared with that of randomly based. Hence we leave the problem open and believe that it deserves further study.
\section{Conclusions and Future work}\label{conclusion}
In this paper, we introduced a semi-supervised community structure detection framework for complex network analysis. The framework adopts a simple strategy to add the supervision of pairwise must-link and cannot-link constraints into the adjacency matrix, which can be regarded as de-noising of the consensus matrix of community structures. The experiments on both the synthetic and real-world networks have demonstrated the effectiveness of the proposed framework. In summary, it can combine the network's functions (background information and domain knowledge) with its topology, making the community structure detection more effective and the results more practical.


We would like to close this paper by raising two interesting problems. Firstly, as we have mentioned above, are there any better methods that can be used for selecting the constraints? A good attempt is the work in \cite{ma2010semi}, which selected the constraints based on various similarity measures, not randomly. Secondly, the proposed framework is very flexible, and can be naturally combined with some other semi-supervised learning models. Researches on this kind of combination are our future working directions.
\section{Acknowledgement}
This work is supported by National Natural Science Foundation of China under Grant No. 61203295. The author is very appreciated the reviewers' valuable comments.

\end{document}